\documentstyle[12pt,epsfig]{article}

\newcommand{\Od}{{\cal O}}

\newcommand{\im}{\mbox{Im}\,}
\newcommand{\re}{\mbox{Re}\,}
\newcommand{\sgn}{\mbox{sgn}}

\newcommand{\be}{\begin{equation}}
\newcommand{\ee}{\end{equation}}
\newcommand{\ba}{\begin{eqnarray}}
\newcommand{\ea}{\end{eqnarray}}

\newcommand{\ar}{\arrowvert}

\newcommand{\ov}{\overline}
\newcommand{\cd}{\! \cdot \!}

\newcommand{\tad}{F_\beta}

\begin{document}

\begin{center}
  {\Large \bf Finite Temperature Pion Scattering to one loop in Chiral
    Perturbation Theory}

\vspace{.5cm} {\large A. G\'omez Nicola$^a$
, F. J. Llanes-Estrada$^a$ and J. R.
Pel\'aez$^{a,b}$.}

\vspace*{.2cm}

\emph{ $^a$Dept. de F\'{\i}sica Te\'orica.
  Univ. Complutense. 28040 Madrid. SPAIN.}

\emph{ $^b$Dip. di Fisica. Univ. degli Studi, Firenze, and INFN,
Sezione di Firenze, Italy}
\end{center}


\vspace{-.5cm}
\noindent
\begin{abstract}
We present the pion-pion elastic scattering amplitude at finite
temperature to one-loop in Chiral Perturbation Theory. The thermal
scattering amplitude properly defined allows to generalize the
perturbative unitarity relation to the $T\neq 0$ case. Our result
provides a model independent prediction of an enhanced pion-pion
low-energy phase shift with the temperature and it has  physical
 applications
within the context of Relativistic Heavy Ion Collisions.
\vspace*{.2cm}

\end{abstract}
\vspace{-.5cm}
\section{Finite Temperature Pion Scattering}
There is a growing interest in the Quark-Gluon-Plasma produced in
Relativistic Heavy Ion Collisions (RHIC) and its evolution into
hadronic matter. After the chiral phase transition, this plasma
cools down rapidly, mainly into a pion gas, which requires  a
description of QCD at temperatures below the chiral phase
transition. With this aim we turn to Chiral Perturbation Theory
(ChPT)
 \cite{we79,gale84} which is
 the low-energy effective theory of QCD. It follows from the identification
of pions as the Nambu-Goldstone bosons associated to the QCD
$SU(2)_L\times SU(2)_R\rightarrow SU(2)_{L+R}$ spontaneous chiral
symmetry breaking and has been already applied to describe
static properties of the low-$T$ pion gas \cite{piongas}. Within ChPT,
 we will analyze the thermal $\pi\pi$ scattering, which
is  relevant for the gas description in terms of transport
equations, for the evolution of the expanding fireball or for
estimating the freeze out time (when the scattering mean time is
larger than the collective expansion time scale). Pion scattering
in the $\sigma$ channel
 has also been related to chiral symmetry restoration \cite{japo1,japo2}.
The thermal amplitude in the vector channel has been studied
within a modified Nambu-Jona Lasinio (NJL) model \cite{NJL1}
explicitly including the $\rho\pi\pi$ vertex and analyzing the
complications due to a non-physical constituent quark-antiquark
threshold near the $\rho$ mass.
 Within  ChPT  only the scattering lengths are available at finite
 $T$
\cite{Kaiser} (for a NJL calculation see \cite{NJL2}).  In this
work we calculate the whole $\pi\pi$ elastic scattering amplitude
at finite temperature to one loop in ChPT, thus valid at  energies
and temperatures well below $\Lambda_\chi\simeq$ 1.2 GeV, both
considered as $\Od(p)$ in the chiral expansion.

A further motivation for our analysis is one of the most striking
observations of RHIC: the modification of the dilepton spectrum
with respect to the vacuum hadronic emission \cite{ceres99}. This
is usually interpreted as an in-medium modification of the $\rho$
mass and width \cite{li95,ele01}. Here we will obtain a completely
model independent and systematic  description of the low energy
tail of this resonance, in terms of $\pi\pi$ scattering. Let us
remark that  we are taking into account the details of the pion
interactions, not only their distribution function \cite{ele01}
whose effect comes from the imaginary part of the amplitude, as we
will see below. The description of the full resonance is beyond
ChPT. Nevertheless, we show here that our amplitude satisfies a
thermal generalization of  perturbative unitarity, which is the
starting point \cite{future} of the successful nonperturbative
unitarization methods used to study resonances like the $\rho$ and
$\sigma$  at $T=0$ \cite{IAM}.

\section{The  $T>0$ Pion Scattering Amplitude } \label{anasec}

\subsection{Definition and interpretation of the thermal amplitude } \label{ana}
Let us start by discussing briefly what we mean by thermal
scattering amplitude in what follows and why it contains
interesting physical information. For our purposes, it will be
enough to assume that the system is in thermal equilibrium and to
neglect finite baryon density effects (as in an ideal RHIC in the
central rapidity region).

In Thermal Field Theory, the word "amplitude" is often used to
denote a thermal Green function rather than an $S$-matrix element,
whose definition in a thermal bath is rather subtle. At $T=0$ the
$S$-matrix element is related to a time-ordered product through
the LSZ reduction formula. Such a time-ordered product can be
calculated at $T\neq 0$, where the time components are ordered
along a complex contour $C$ \cite{lebellac}. In the imaginary-time
formalism (ITF) $C=[0,-i\beta]$ so that the time arguments become
Euclidean $\tau=it \in C$, with $\beta=T^{-1}$ and $t$ being the
Minkowski time. The Feynman rules in the ITF are simply those at
$T=0$ but replacing all the zeroth components of momenta by a
discrete set of frequencies $q^0\rightarrow i\omega_n=2\pi i n T$
and the loop integrals  by Matsubara sums, i.e., $ (2\pi)^{-1}\int
dq^0 \rightarrow i T \sum_{n=-\infty}^{\infty}$. Since the
vertices remain unchanged the set of diagrams is exactly the same
as for $T=0$. The ITF is  appropriate to calculate thermodynamical
quantities, but it can also be used to analyze dynamical processes
with real energies. In fact,  carrying out the Matsubara sums
first, the ITF Green functions
 yield analytic functions of $i\omega_n$
 everywhere {\it off} the real axis \cite{lebellac}.
 Therefore, they can be continued analytically from
discrete external complex to continuous real energies $E$, usually
with the prescription $i\omega_n\rightarrow E +i\epsilon$. It is
worth mentioning that the thermal Green functions thus obtained
correspond to the {\it retarded} functions in the real-time
formalism (RTF) \cite{bani94,kobes} for which $C$ includes the
real axis. One could also calculate the time-ordered product
directly in the RTF, where the external energies remain real, but
the number of fields is doubled \cite{lebellac}. In this sense,
let us remark that in the RTF it has been shown \cite{kobes} that
the $T\neq 0$ retarded functions have a causal and analytic
structure which allows for a description in terms of dispersion
relations, unlike the $T\neq 0$ RTF time-ordered product. Hence,
the retarded functions  are suitable candidates to generalize
important properties of the $T=0$ amplitude such as unitarity, as
we will see below.

Therefore, back to  our present case, consider  the $T=0$
$S$-matrix element corresponding to $\pi\pi$ scattering, written
through the LSZ formula in terms of the Green function of the
time-ordered product of four pion fields. The thermal scattering
amplitude we will consider is {\it defined} with respect to such $T=0$
amplitude by keeping the pion initial and final asymptotic states
as those for $T=0$ but  i) replacing the Green function by a
thermal expectation value in the ITF   and ii) performing the
usual analytical continuation into continuous real energies. As
discussed above, this corresponds to the {\it retarded} four-point
function in the RTF and, as we will see, the thermal amplitude
thus defined has the correct analytic structure, allowing for a
suitable extension of the perturbative unitarity at $T\neq 0$. Let
us remark that the same definition for the pion scattering thermal
amplitude has been used in \cite{NJL1,Kaiser,NJL2}.

An alternative point of view would have been to consider the RTF
thermal reaction rate ${\cal R}$  for the {\it inclusive} reaction
$\pi\pi$+Heat Bath $\rightarrow$ $\pi\pi$ + anything
\cite{niegawa} defined as the thermal average of the transition
probability $S^*S$, with the $T=0$ $S$-matrix element between an
initial and final asymptotic state {\it including}  the heat bath
collective states. We remark that this is a different object than
the thermal amplitude we consider here, thus providing information
on different physical observables. Thus, while ${\cal R}$ can be
interpreted as an "average of probability" when the {\it free}
heat bath is taken also as an asymptotic state, the thermal
amplitude we consider here can be viewed rather as (the retarded
part of) an "average of amplitude" of the probability that
 two initial free pions scatter
 in the interacting thermal bath and two final free pions leave it.
 In principle, there is no obvious relation between the two,
 except for the case of two-point functions, where the imaginary
 part of the retarded RTF Green function is proportional to the
 difference between the direct decay rate, given by ${\cal R}$  and
 the inverse one, related to ${\cal R}$ by detailed balance
 \cite{niegawa,kobes}.
 A detailed calculation of the $T\neq 0$ reaction rate for the pion
scattering using the rules given in \cite{niegawa} would be very
interesting but it lies beyond the scope of this work.

\subsection{The one loop ChPT calculation}

The  $N_f=2$ chiral lagrangian at  $\Od(p^2)$
 is that of the $O(4)$ nonlinear sigma model:
\begin{equation} \label{nlsm}
{\cal L}_2=\frac{f^2}{2}\left( \partial_\mu U^T
\partial^\mu U +2\chi^T U \right)
\end{equation}
where $f$ is the $T=0$ pion decay constant in the chiral limit,
$U(x)\equiv \left(\sqrt{1-\pi^2/f^2},\vec{\pi}/f \right)$ is an
$O(4)$ vector satisfying $U^T U=1$, with $\pi_a$ ($a=1,2,3$) the
pion fields and $\chi=\left(m^2,\vec{0}\right)$. Here, $m$ is the
pion mass to lowest order (in the isospin limit). For simplicity,
we do not show external sources, which are nevertheless needed to
obtain the  $\Od (p^4)$ relations between the physical $m_\pi$,
$f_\pi$ and $m$, $f$ \cite{gale84}. The $\Od (p^4)$ lagrangian
${\cal L}_4$ includes seven independent terms multiplied by
low-energy constants, called $\bar l_i$ after renormalization.

There is another subtlety to be borne in mind. The loss of Lorentz
covariance imposed by the choice of the thermal bath rest frame
implies that we have to separate explicitly the time and
space-like components of any four-vector. In particular, for an
elastic two-body collision with four-momenta $k_1  k_2\rightarrow
k_3 k_4$, if we call $ {\bf S}=k_1+k_2$, $ {\bf T}=k_1-k_3$ (not
to be confused with the temperature) and $ {\bf U}=k_1-k_4$, the
amplitude will depend differently on $ \bf{S_0}$, $\vert \vec{
\bf{S}}\vert$, $ \bf{T_0}$, $\vert \vec{ {\bf T}}\vert$, $
\bf{U_0}$ and $\vert \vec{ {\bf U}}\vert$. In particular, crossing
symmetry on the Mandelstam variables $s={\bf S^2},t={\bf
T^2},u={\bf U^2}$ no longer holds. Let us recall that at $T=0$ any
elastic $\pi\pi$ amplitude is related to that of
$\pi^+\pi^-\rightarrow\pi^0\pi^0$, called  $A(s,t,u)$, by isospin
and crossing transformations. Nevertheless, since the temperature
does not modify the vertices, crossing symmetry still holds in
terms of the $S$, $T$ and $U$ {\it four-momenta}, and we can still
write any $\pi\pi$ thermal amplitude in terms of  the thermal
$A({\bf S},{\bf T},{\bf U};\beta)$: \ba A({\bf S},{\bf T},{\bf
U};\beta)=A_2 ({\bf S},{\bf T},{\bf U}) + A_4^{pol}({\bf S},{\bf
T},{\bf U}) \label{Aparts}+ A_4^{tad} ({\bf S},{\bf T},{\bf
U};\beta)+A_4^{uni} ({\bf S},{\bf T},{\bf U}; \beta), \ea where we
have followed the usual $T=0$ notation: $A_2$ is  the ${\cal L}_2$
tree level contribution and $A_4^{pol}$ contains the ${\cal L}_4$
tree level, depending on the  $\bar l_i$, plus those  polynomials
coming from the renormalization of the loop integrals. Both $A_2$
and $A_4^{pol}$ are temperature independent.  The $A_4^{uni}$ term
represents those contributions from loops with two propagators
proportional to the  three {\em independent} integrals $J_0,J_1$
and $J_2$ defined in eq.(\ref{defJ}) which yield the correct
analytic structure at $T=0$ and will ensure perturbative unitarity
in all channels. The
 terms  collected in $A_4^{tad}$ are  proportional
to the tadpole integral $\tad$ in eq.(\ref{deftad}) and
include all tadpole diagrams as well as terms coming from  loop
integrals  (see Appendix \ref{ap:loopint}).

Taking these remarks into account, the thermal
amplitude is obtained by adding
to the temperature independent
part in \cite{gale84} the following temperature dependent corrections:
\ba \label{a4tad}
%
&&\Delta A_4^{tad} ({\bf S},{\bf T},{\bf U};\beta)= \frac{\Delta\tad}{f^4}\left\{
-s+t+u+\frac{5 m^2}{2}  - \left[ \frac{2 {\bf T_0}}{\ar \vec{{\bf T}} \ar^2}
(k_1^0 \vec{k}_2 + k_2^0 \vec{k}_1)\cd \vec{{\bf T}}+2
\left(k_1-k_2\right) \cdot {\bf T}  \right.\right.  \nonumber\\
%
&&\hspace{2cm}+\frac{2(\vec{k_1}\cdot \vec{{\bf T}})(\vec{k_2}\cdot \vec{{\bf T}})}{
\vert\vec{{\bf T}}
\vert^4}\left[3\left(\frac{t}{2}+{\bf T_0}^2\right)+\vert\vec{{\bf T}}
\vert^2\right] +\left.\left.\frac{t(\vec{k_1}\cdot\vec{k_2})}{
\vert\vec{{\bf T}}\vert^2}+\left({\bf T}\leftrightarrow
{\bf U}\right)\right]\right\}
\ea
\ba \label{a4uni}
%
&&\hspace*{-1cm}\Delta A_4^{uni} ({\bf S},{\bf T},{\bf U};\beta)= \nonumber\\
&&\hspace*{-1cm}\frac{1}{f^4} \left\{
\left(s-m^2\right)\left[- 2{k_1}^0 \Delta J_1({\bf S})\right.
+\frac{\Delta J_0({\bf S})}{2}\left(m^2+3s-2k_3\cd {\bf S}\right)
-\frac{\vec{k}_1 \cd \vec{{\bf S}}(s\Delta J_0({\bf S}) -2{\bf S_0} \Delta
J_1({\bf S}))}{\ar\vec{{\bf S}} \ar ^2} \right]
\nonumber\\ &&+ \left[-\frac{m^2}{\ar \vec{{\bf T}} \ar^2}\left[
\left(\vec{k}_1-\vec{k}_2\right)\cd \vec{{\bf T}} \left[t \Delta
J_0({\bf T})-2{\bf T_0} \Delta J_1({\bf T})\right] \right]+m^4 \Delta J_0({\bf T}) -2
(k_1^0-k_2^0) m^2 \Delta J_1({\bf T}) \right.
 \nonumber\\
&&-4
k_1^0 k_2^0 \Delta J_2({\bf T}) -\frac{2}{\ar \vec{{\bf T}} \ar^2} \left(k_1^0
\vec{k}_2 + k_2^0 \vec{k}_1\right)\cd \vec{{\bf T}} \left[t \Delta
J_1({\bf T}) - 2 {\bf T_0} \Delta
J_2({\bf T}) \right]\nonumber\\
&&- \frac{2(\vec{k_1}\cdot \vec{{\bf T}})(\vec{k_2}\cdot \vec{{\bf T}})}{
\vert\vec{{\bf T}} \vert^4}\left[\left(3{\bf T_0}^2-\vert\vec{{\bf T}}
\vert^2\right)\Delta J_2 ({\bf T})\right. -\left.3{\bf T_0} t \Delta J_1
({\bf T})+\left(\frac{3t^2}{4}+m^2\vert\vec{{\bf T}} \vert^2\right)\Delta J_0
({\bf T})\right]
\nonumber\\
&&+\frac{2\vec{k_1}\cdot\vec{k_2}}{
\vert\vec{{\bf T}}\vert^2}\left[t\Delta J_2({\bf T})-t {\bf T_0} \Delta J_1
({\bf T})\right. +\left.\left.\left.\left(\frac{t^2}{4}+m^2 \vert\vec{{\bf T}}
\vert^2\right) \Delta J_0({\bf T})\right] +\left({\bf T}\leftrightarrow
{\bf U}\right)\right]\right\}
 \ea

Note that the above equations in Minkowski space-time are obtained
from the Euclidean expressions by replacing $k^0\rightarrow -i E$
for all Euclidean momenta $k_0$, with $E$ real, and analytically
continuing the $J_k$ as showed in eq.(\ref{eutomin}). Note also
that, for clarity, we have not simplified the $k_i$ four-momenta
in terms of ${\bf S},{\bf T}$ and ${\bf U}$. In addition, the
$T$-corrections to $m_\pi$ and $f_\pi$ \cite{piongas} are included
in $\Delta A_4^{tad}$, consistently with our definition
 of the thermal amplitude.
 Recall that the $T=0$ physical $f_\pi$ and $m_\pi$ are
 \cite{gale84}:
\begin{eqnarray} \label{mpi}
m_\pi^2 = m^2\left(1 - \frac{\ov{l}_3 m^2}{32 \pi^2 f^2} \right)
\quad , \quad f_\pi = f\left(1 + \frac{m^2}{f^2} \frac{\bar
l_4}{16 \pi^2}\right).
\end{eqnarray}

 We have performed three different consistency checks on the
thermal amplitude: First, we have recovered the Lorentz invariant
$T\rightarrow 0^+$ limit of \cite{gale84}. Since all the
temperature dependence is on the tadpoles and $J_k$ functions, we
are thus checking their  coefficients, since our calculation was
carried out without  Lorentz covariance. The other two checks
concern partial waves and the c.o.m. frame, i.e, $\vec{{\bf S}}=0$
or pions at rest with the thermal bath. Customarily, the $\pi\pi$
scattering amplitude is projected into partial waves $a_{IJ}$ of
definite isospin $I$ and angular momentum $J$, and the standard
definitions \cite{gale84,IAM} are still applicable at $T\neq0$,
but with crossing symmetry  in terms of the ${\bf S},{\bf T}$ and
${\bf U}$ four-vectors. This procedure is equivalent to a direct
calculation of the isospin amplitudes without using the $A$
function, as done in \cite{Kaiser}, whose analytic expressions for
the thermal scattering lengths we have reobtained, thus checking
the real part of our partial waves with an independent
calculation. The third check concerns the imaginary parts of the
thermal partial waves in the different channels. This requires a
generalization of the perturbative unitarity relation that we
present next, which, in turn, will provide a natural physical
interpretation of the imaginary part of our thermal amplitude in
terms of direct (emission) and inverse (absorption) processes in
the thermal bath.

\subsection{Thermal unitarity}
\label{sectu}

For simplicity,  we drop the $IJ$ indices in the following. At
$T=0$ unitarity constraints the partial waves, for $s>4m_\pi^2$
and below other inelastic thresholds, to satisfy
\begin{equation}  \label{unit0}
 \im a(s)=\sigma (s) \vert a (s) \vert^2
\ , \quad \sigma (s)=\sqrt{1-4 m_\pi^2/s} \ ,
\end{equation}
whereas the ChPT series only satisfies unitarity perturbatively, i.e.,
\begin{equation}  \label{pertunit0}
 \im a_2 (s)= 0, \quad \im a_4 (s) = \sigma (s)
\left\vert a_2(s) \right\vert^2, \quad  ....
\end{equation}

We now generalize the above {\it perturbative} relation for any
one-loop calculation, including those within ChPT. Let us recall
that for one-loop elastic amplitudes, any imaginary part should
come from loop diagrams with two propagators and in particular
from the one-loop integrals $J_k$ ($k=0,1,2$) given in
eq.(\ref{defJ}). As an illustration we will analyze $J_0$, since
the other cases are completely analogous. We start then by
performing the frequency sums in the standard fashion
\cite{lebellac}: \ba \label{j0gen} J_0 (\omega,\vert
\vec{Q}\vert;T)= -\int \frac{d^{D-1}q}{(2\pi)^{D-1}}
\sum_{s_1,s_2=\pm 1} \frac{s_1 s_2}{4 E_q E_{q-Q}} \frac{1+n
\left(s_1 E_q\right) + n\left(s_2 E_{q-Q}\right)}{\omega-s_1
E_q-s_2 E_{q-Q}} \ea
 where $E_k^2=\vert\vec{k} \vert^2+m^2$,
$n(x)=\left(e^{\beta x}-1\right)^{-1}$ is the Bose-Einstein
distribution function and $\omega\equiv iQ_0=2\pi k T i$ with $k$
integer, is the external frequency. Note that eq.(\ref{j0gen})
provides the analytic continuation of $J_0 (\omega)$  for $\omega$
off the real axis. Taking now $\im J_0 (E+i\epsilon)$ with $E$
real (see our previous discussion on the thermal amplitude) the
two $s_1=-s_2$ terms cancel each other. The two cases
$s_1=s_2=\pm1$ require $E\pm E_q\pm E_{q-Q}=0$ which imply $\pm
E-\sqrt{\vert \vec{Q}\vert^2+4m^2}>0$, respectively. Summarizing:
\ba \label{thphsp} \im J_0 (E+i\epsilon,\vert \vec{Q}\vert;T)&=&
\pi\,\sgn
(E)\theta (s-4m^2) \\
&\times&\int \frac{d^3\vec{q}}{(2\pi)^3} \frac{1}{4 E_q E_{q-Q}}
\left[1+n \left(E_q\right) + n\left(E_{q-Q}\right)\right] \delta
\left(\vert E \vert- E_q- E_{q-Q}\right)\nonumber \ea where
$s=E^2-\vert \vec{Q}\vert^2$ and $E$ real. Hence $J_0 (s)$ has a
cut on the real axis starting at $s=4m^2$, as in the $T=0$ case.
Our analysis is similar to that of a decaying particle in medium
\cite{Weldon}, where the integral in eq.(\ref{j0gen}) is the
thermal self-energy.  In fact, the interpretation of
eq.(\ref{thphsp}) follows from the observation that $ 1+n (E_1)+n
(E_2) =[1+n (E_1)][1+n (E_2)]-n (E_1) n (E_2)$. Therefore, in our
thermal amplitude and  for positive energy, the in-medium pions
increase the available phase space by $(1+n_1)(1+n_2)$
(corresponding to the thermal enhancement of the two pion
production) and decrease it by $n_1 n_2$ (when the two initial
pions collide with thermal pions and are absorbed) analogously to
the  stimulated emission and absorption processes discussed in
\cite{Weldon}.  Alternatively, we could have calculated
in the RTF, arriving to the same result for the retarded
self-energy, unlike the  RTF time-ordered product, which  yields
the {\it sum} of the emission and absorption terms rather than
their difference \cite{kobes}.

Partial waves are defined in the c.o.m. frame, where we
start studying the $S$-channel
diagrams, setting $\vec Q=0$ in eq.(\ref{thphsp}). The
generalization to one-loop ChPT is straightforward. Since ${\cal
L}_2$ contains only two derivatives, one needs loop integrals with
up to four momenta in the numerator, whose imaginary part can be
related to $J_0$, $J_1$ and $J_2$ in eq.(\ref{defJ}) (see Appendix
\ref{ap:loopint}). The $J_{1,2}$ analysis is similar to that for
$J_0$, and we find
\begin{eqnarray} \label{prevunigen}
\im J_k (E+i\epsilon, \vec{0};T)&=&\frac{\sgn (E)}{16\pi} \theta
(s-4m^2) \sigma_T (E) \left(\frac{E}{2}\right)^k, \quad k=0,1,2\\
\label{sigmaT}
\hbox{where} \quad \sigma_T (E)&=&\sigma(E^2)\left[1+\frac{2}{\exp({\beta\vert E
\vert/2})-1}\right]
\end{eqnarray}
is the thermal two-particle phase space in the c.o.m. frame
\cite{Weldon}. Note that in (\ref{prevunigen}) we have continued
analytically the $J_k$ functions to real energies as indicated in
eq.(\ref{eutomin}) .

Concerning the  $T$ and $U$ channels, in the c.o.m. frame and
 for $s>4m_\pi^2$ we have to evaluate $J_k(0,\vert \vec{{\bf T}} \vert) $
 (see Appendix \ref{ap:loopint}) which is real, according to
 eq.(\ref{thphsp}).
Therefore, the same $T=0$ derivation of eq.(\ref{pertunit0})
follows for $T>0$ by replacing $\sigma\rightarrow \sigma_T$, i.e.:
\begin{equation}
\label{pertunitT}
\im a_2 (s)= 0, \qquad \im a_4 (s;\beta) = \sigma_T ({\bf S_0})
\left\vert a_2(s)\right\vert^2\ , \quad {\bf S_0}>2m_\pi
\end{equation}
This generalizes the perturbative unitarity relation to $T\neq 0$,
providing a neat physical interpretation of the imaginary part of
the thermal amplitude in terms of emission and absorption of pion
pairs in the thermal bath. Note  that the one-loop thermal
amplitude considered here, in terms of retarded propagators, has
the same unitarity cut as the $T=0$ one. In fact, it can be shown
 that they have the same analytic structure in the
complex plane \cite{future},
 which gives support to the idea that the thermal retarded
 functions provide the natural extension of amplitudes as far as
  their causal and analytic properties are concerned.
  We remark that our {\em perturbative} result (\ref{pertunitT}) is
consistent within ChPT, where the external energies and $T$ remain
small  and so do the Bose-Einstein factors. For higher orders one
could have intermediate states in the thermal bath with more than
two pions, weighted by higher powers of $n$. We have explicitly
checked that our final result for the amplitude satisfies
eq.(\ref{pertunitT}) for the $IJ$=00,11,$20$ channels.

 Finally, let us note that  regarding the $\pi\pi$ scattering
 in the $11$ channel as the exchange of a  $\rho$ resonance
 \cite{Res}, the pion scattering amplitude would be proportional to
 the $\rho$ thermal retarded self-energy, whose imaginary part to
 one loop is proportional to $\sigma_T/\sigma$ and gives the  net
 decay rate of the $\rho$ in the thermal bath. Note
 that, from this point of view, the stimulated
emission contribution, proportional to
 $\left[1+n(\vert E \vert/2)\right]^2$,  would give the reaction
rate discussed in
 \cite{niegawa}
 for the  decay of a $\rho$ at rest into two pions. Hence, we expect
 to be able to extract physical information
 about the thermal properties of physical resonances arising in
 our $\pi\pi$ scattering amplitude, which we have
calculated without introducing those resonances
  as explicit degrees of freedom. This is indeed the case, as shown recently
in \cite{future}, where the thermal mass and width of the $\rho$ and $\sigma$
   are
 read off  from the complex poles (in the second Riemann sheet)
 of the thermal nonperturbative  amplitude,
 obtained from the ChPT one by unitarization. At low $T$, the
 $\rho$ width grows with the $\sigma_T/\sigma$ factor, as
our previous discussion
  suggested.  This
 provides a relevant physical application of our present analysis, which
is totally model independent, in contrast with the very recent results
in  \cite{future}, which maybe more direct to extract information
on resonances but relies on the choice of unitarization method.

\section{Discussion and conclusions}
\label{conc} 
\begin{figure}[h]
\hspace*{-1cm} \hbox{\psfig{file=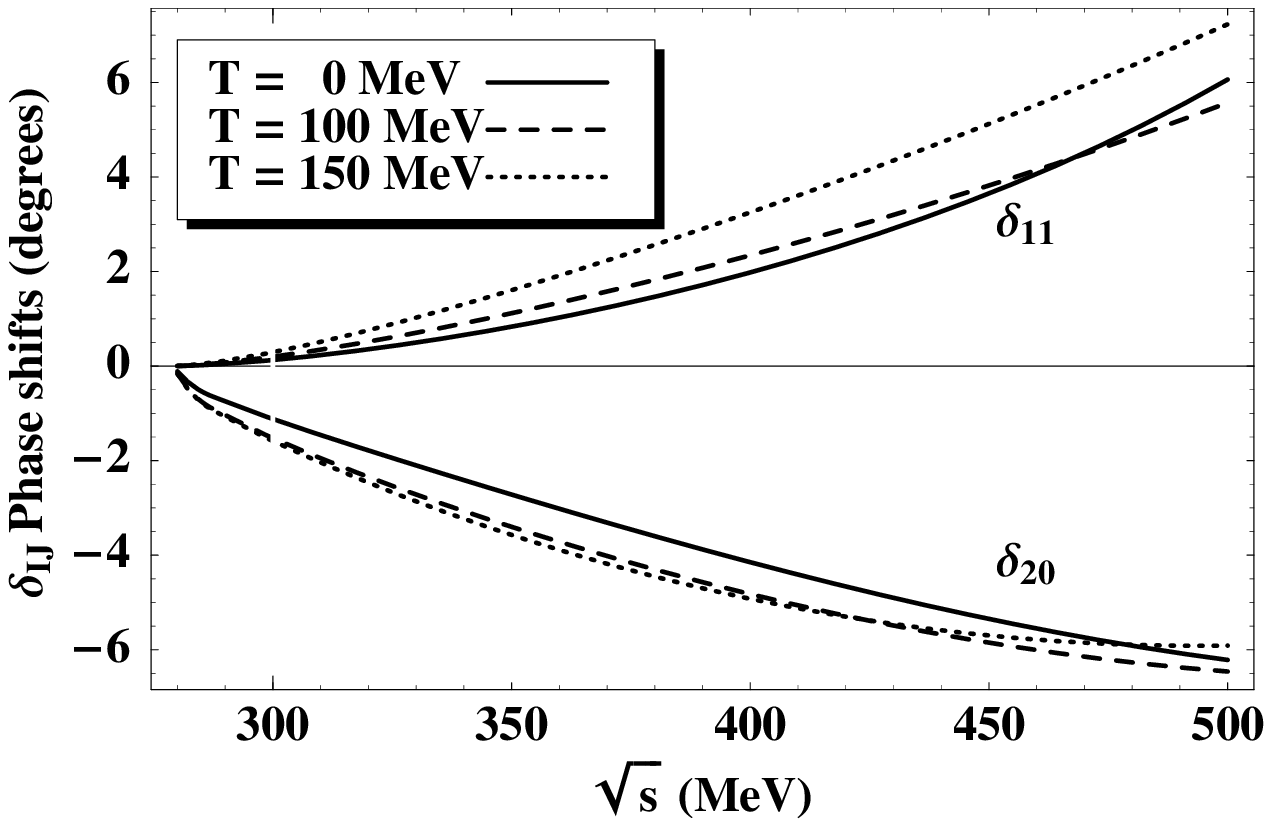,width=8.5cm}
\psfig{file=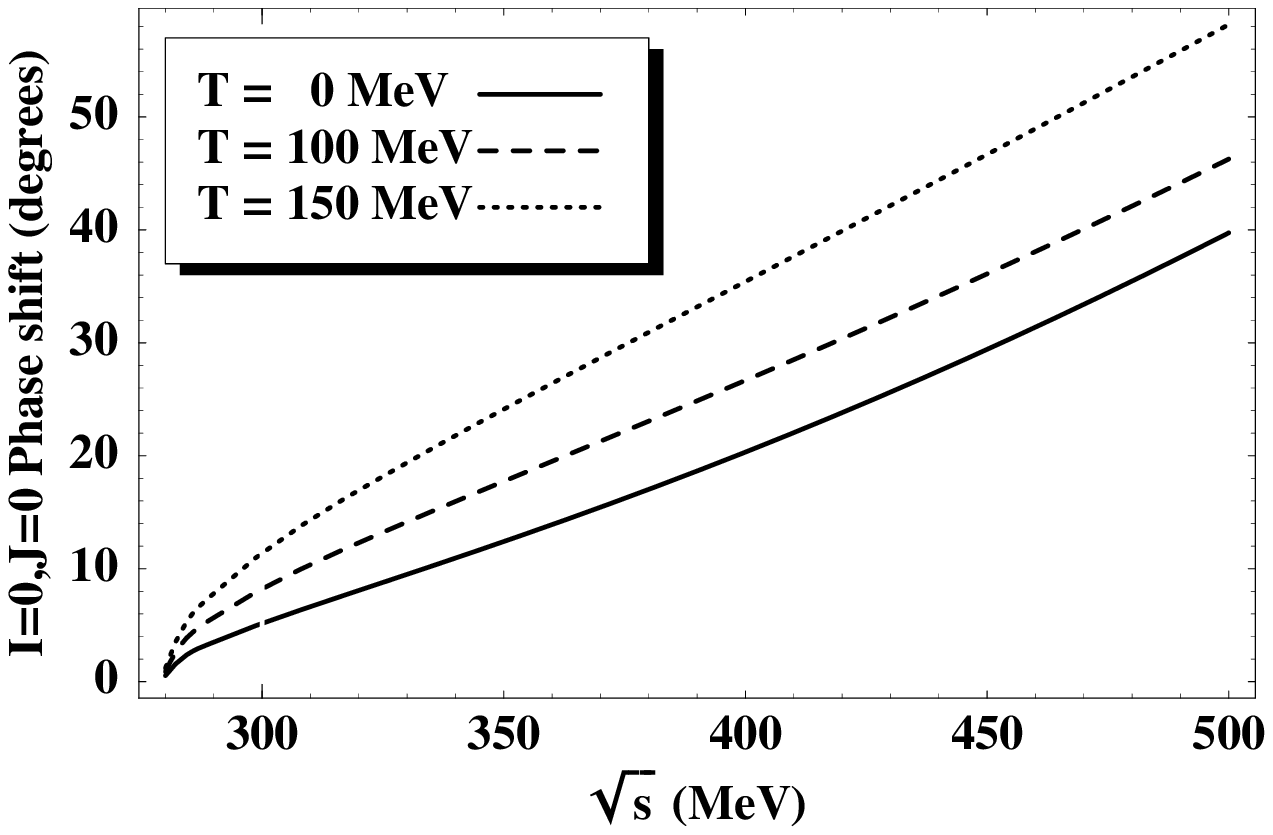,width=8.5cm}} \caption{\rm
\label{fig:diagrams} Temperature evolution of the phase shifts
$\delta_{IJ}$ for  $IJ=11,00,20$. }
\end{figure}

From our previous results, we have also calculated
the scattering phase shifts, $\delta_{IJ}$ in different channels.
This is a simple and easy way of representing the complex
amplitude in terms of just one parameter, since elastic unitarity
implies $a= \exp(\delta)\sin(\delta)/\sigma_T$. Since, for a given
energy, the thermal phase space is fixed, the ``strength of the
interaction'', i.e. the modulus of the amplitude, is given only in
terms of $\sin \delta$ This is similar to the $T=0$ case, and
perturbatively corresponds to $\delta_T\simeq \sigma_T
\left(\sqrt{s}\right)\left[a_2(s)+\re a_4(s,\beta)\right]$ . The
results are plotted in Fig.\ref{fig:diagrams} up to 500 MeV,
roughly where ChPT is reliable. Note that the absolute value of
all the phase shifts is increased, but each channel keeps its
attractive or repulsive nature (the sign of the phase shifts). The
phase shift increase we find is basically due to the Bose
enhancement factor in $\sigma_T$, the real part and the modulus of
the perturbative amplitude changing very slightly with $T$ at low
energies. In fact, in the  $IJ=20$ channel, $\vert \re a_4 \vert$
{\em decreases} with $T$ for low energies (see \cite{Kaiser}). A
strong enhancement of $\vert a_{00} \vert^2$ near threshold would
be a signal of chiral symmetry restoration \cite{japo1,japo2} but
we do not observe such effect here. We also remark that not all
the thermal effects in our one-loop calculation can be accounted
for with a thermal redefinition of $f_\pi$ \cite{japo2} since in
that case, with the decreasing  $f_\pi (T)$ given in
\cite{piongas}, we would see a larger increase of $\vert a\vert^2$
in {\it all} channels. Note also that our result can be
alternatively interpreted as an enhanced effective pion
interaction  if we were describing them
 with the $T=0$ expressions, although bearing in mind that
 the main contribution comes from the phase space factor.

The $IJ=11$ channel is particularly interesting, since it is
dominated by the $\rho$ resonance, whose thermal deformation has
been related to the anomalous dilepton spectra in RHIC
\cite{li95,ele01}. By analyzing this channel at  low energies, we
are  studying the tail of the resonance. The low-energy
temperature enhancement of the phase shift can then be interpreted
as a stronger thermal effect of the rho, although with pure ChPT
calculations we cannot disentangle whether it is due to a larger
width or to a smaller mass. Qualitatively, the chiral combination
of parameters that dominate the phase shift in this channel is
$\bar l_1-\bar l_2\sim \Gamma_\rho f^4/M_\rho^5$, following the
resonance saturation hypothesis \cite{gale84,Res}. In order to
obtain larger phase shifts at $T=0$, we need  somewhat larger
$\bar l_1-\bar l_2$ values. Thus, the {\it low-energy} thermal
phase shifts would correspond at $T=0$ to a larger $\Gamma_\rho
f^4/M_\rho^5$ ratio, although with the $T=0$ part only we cannot
mimic accurately the {\it low-energy} thermal behavior. Therefore,
our results are consistent with most models that expect a low-$T$
sizable thermal increase of the rho width due to the thermal
increase of phase space, with an almost constant thermal $M_\rho$
\cite{dompis,ele01}. Our model independent
 perturbative result is confirmed by
 our  discussion at the end of section \ref{sectu} and by
 the unitarized results in \cite{future}.

In conclusion, we have calculated the
$\pi\pi$ scattering to one loop in Chiral Perturbation Theory at
finite temperature $T$. Our thermal amplitude has been calculated
with $T=0$ asymptotic states and  the thermal  ITF Green function,
whose analytic continuation  corresponds to
 the retarded RTF Green function, which has
a suitable causal and analytic structure.
 Despite the loss of Lorentz covariance of the thermal formalism,
any $\pi\pi$ elastic amplitude can still be related to a single
one by means of a generalized crossing symmetry in terms of the
external four-momenta. We have found the thermal generalization of
perturbative partial wave unitarity to one loop, which
amounts to replacing the $T=0$ two-particle phase space with its
thermal version. This accounts for emission and absorption of pion
pairs inside the thermal bath, providing a physical interpretation
of the imaginary part of the thermal amplitude and giving the
expected analytic structure  above threshold, consistently with
causality and thermal unitarity. We have checked that our
amplitudes satisfy this constraint. With these amplitudes, we have
studied the low energy behavior of thermal $\pi\pi$ elastic
scattering in a model independent formalism, finding an enhanced
phase shift in all channels, as dictated by the thermal phase
space, but with little change in the modulus of the partial waves.
Physically, this means that one cannot mimic the low-$T$ behaviour
of dynamical quantities just by scaling $f_\pi\rightarrow f_\pi
(T)$.

One of the  main physical observables that can be related to our
thermal scattering amplitude are the masses and widths of the
$\rho$ and $\sigma$ mesons \cite{future}. We have given here a
low-$T$ qualitative description for the case of the $\rho$.
Further quantitative and detailed information can be obtained if
these results are extended to higher energies beyond pure ChPT. In
particular, by means of unitarization methods \cite{IAM} that
generate dynamically light resonances like the $\rho$ and $\sigma$
mesons \cite{future}. Hence, our results are useful as a starting
point for the description of relevant physical processes where
chiral symmetry and unitarity play a crucial role, such as the
properties of those resonances in the plasma formed after a
Relativistic Heavy Ion Collision.

\subsection*{Acknowledgments}
 The authors thank A. Dobado for useful comments and
discussions. F. J. Ll-E thanks P. Bicudo, S. Cotanch, P. Maris, E. Ribeiro
and A. Szczepaniak too. Work supported from the Spanish
CICYT projects,  FPA2000-0956,PB98-0782 and BFM2000-1326.
\appendix
\section{Loop integrals} \label{ap:loopint}
The following are the integrals needed for our calculation in an
arbitrary reference frame (not necessarily the c.o.m frame,  where
pions are at rest with the thermal bath). The signature of the
metric in Euclidean space is $(- \ - \ - \ -)$ and we will use the
notation
$$\int_\beta \frac{d^Dq}{(2\pi)^D}
\equiv \frac{1}{\beta} \sum_{n=-\infty}^{n=+\infty} \int
\frac{d^{D-1}q}{(2\pi)^{D-1}}$$
where $\beta=1/T$, $q_0=2\pi n T$
and  $Q_0=2\pi k T$ with $k$ integer denotes an external
frequency. Also we denote $\Delta G(T)\equiv G(T)-G(T=0)$.
First, we have the tadpole integral:
\begin{equation} \label{deftad}
\tad \equiv \int_\beta \frac{d^Dq}{(2\pi)^D} \frac{1}{q^2-m^2} \ .
\end{equation}
Then, three integrals of pion loops with an external momentum
flow: \ba J_k (Q_0,\vert \vec{Q}\vert;T)&\equiv& \int_\beta
\frac{d^Dq}{(2\pi)^D}\frac{q_0^k}{(q^2-m^2) ((q-Q)^2-m^2)} \ ,
 \quad k=0,1,2 \label{defJ} \ea All other  integrals with momenta
in the numerator can be reduced to those four as
\ba
\int_\beta \frac{d^Dq}{(2\pi)^D}\frac{q_i}{(q^2-m^2)
((q-Q)^2-m^2)} &=& -\frac{Q_i}{\ar \vec{Q}\ar^2}
\left[ Q_0 J_1 + \frac{1}{2}J_0
Q^2 \right]        \ .\\
\int_\beta \frac{d^Dq}{(2\pi)^D}\frac{q_\mu Q^\mu}{(q^2-m^2)
((q-Q)^2-m^2)} &=& \frac{Q^2}{2} J_0  \ .\
\ea
\ba
\int_\beta \frac{d^Dq}{(2\pi)^D}\frac{q_0 q_i}{(q^2-m^2)
((q-Q)^2-m^2)} &=& -\frac{Q_i}{\ar \vec{Q}\ar^2} \left[
Q_0 J_2+ \frac{Q^2}{2}J_1 + \frac{Q_0}{2} \tad \right] \ .
\\
\int_\beta \frac{d^Dq}{(2\pi)^D}\frac{q_i q_j}{(q^2-m^2)
((q-Q)^2-m^2)} &=& Q_i Q_j I_{2a} +g_{ij} I_{2b}
\ea
\ba
\int_\beta \frac{d^Dq}{(2\pi)^D}\frac{q_0 q\cd q}{(q^2-m^2)
((q-Q)^2-m^2)}&=& Q_0 \tad + m^2 J_1
\\
\int_\beta \frac{d^Dq}{(2\pi)^D}\frac{q_i q\cd q}{(q^2-m^2)
((q-Q)^2-m^2)} &=& Q_i \tad- m^2 \frac{Q_i}{\ar
\vec{Q}\ar^2} \left( Q_0 J_1 + \frac{Q^2}{2} J_0 \right) \ .
\ea
\ba
\int_\beta \frac{d^Dq}{(2\pi)^D}\frac{(q\cd q)^2}{(q^2-m^2)
((q-Q)^2-m^2)}= (Q^2 +2m^2) \tad + m^4 J_0 \ .
\ea
where \ba \label{I2aI2b} I_{2a}(Q)&=&
\frac{1}{(D-2)\ar\vec{Q}\ar^4} \left[ \left( (D-1) Q_0^2 +
\ar\vec{Q}\ar^2\right)J_2 + (D-1)Q_0 Q^2 J_1 + \left(
\frac{D-1}{4}Q^4 +m^2 \ar\vec{Q}\ar^2 \right)J_0  \right. \nonumber\\
  &+&\left.\left( (\frac{Q^2}{2}+Q_0^2)(D-1)+ \ar\vec{Q}\ar^2
\right) \tad \right] \nonumber\\
I_{2b}(Q)&=& \frac{1}{(D-2)\ar\vec{Q}\ar^2} \left[ -Q^2 J_2 +Q_0
Q^2 J_1 + \left( \frac{Q^4}{4} +m^2 \ar\vec{Q}\ar^2 \right)J_0
-\frac{Q^2}{2} \tad \right] \ .
\end{eqnarray}

The above integrals have the appropriate $T\longrightarrow 0$,
$m\longrightarrow 0$ limits and at $T=0$ they can be written just
in terms of the tadpole $F$ and $J_0$ \cite{gale84}. Recall that
all the UV divergences of the $\tad$ and $J$ integrals are
contained in the $T=0$ part, since $\Delta F$ and $\Delta J_k$
always contain Bose-Einstein functions vanishing exponentially for
large momentum. Therefore, using dimensional regularization, we
always separate the $T=0$ part, regularized and renormalized as in
\cite{gale84} and set $D=4$ in $\Delta\tad$ and $\Delta J_k$.

The functions $J_k$ in eq.(\ref{defJ}) can be analytically
continued after performing the Matsubara sums. All the Euclidean
discrete frequencies are then continued to $-iE$ with $E$ real and
continuous, i.e, \ba J_k(Q_0,\vert \vec{Q}\vert;T)&\rightarrow&
(-i)^k \,J_k( -iE,\vert \vec{Q}\vert;T) \label{eutomin} \ea In the
c.o.m. frame the imaginary part of the $J$ integrals, can be
calculated exactly, e.g, eq.(\ref{prevunigen}). When the c.o.m
frame $\vert \vec{{\bf S}} \vert={\bf T_0}={\bf U_0}=0$, coincides
with the thermal bath rest frame, only $\tad$, $J_0({\bf S})$,
$J_0({\bf T})$, $J_2({\bf T})$ have to be evaluated (note that
$J_0({\bf U})$, $J_2({\bf U})$ can be computed from the
corresponding ${\bf T}$ integrals). Specifically
\begin{eqnarray}
J_2({\bf S})&=&\frac{{\bf S_0}^2}{4}J_0({\bf S})+\frac{\tad}{2}\ , \quad J_1({\bf S})=
\frac{{\bf S_0}}{2} J_0({\bf S}) \ , \quad J_1({\bf T})=0
\end{eqnarray}
The real part of $\Delta J_0$ analytically continued as in
eq.(\ref{eutomin}) in the $S$-channel reads
\begin{equation}
\re \Delta J_0({\bf S}) = -\frac{1}{\pi^2} PV \int_m^\infty
 dE \ \frac{ q(E) n(E)}{s-4E^2}
\end{equation}
where $q(E)=\sqrt{E^2-m^2}$ $n(E)=\left(e^{\beta E}
-1\right)^{-1}$ and $PV$ denotes Cauchy's Principal Value. Note
that the above expression is valid for $\omega$ real
($s=\omega^2$) and the pole of the integrand is only present for
$s>4m^2$, where $\im J_0\neq 0$.  Finally, in the $T (U)$
channels, we get
\begin{eqnarray*}
\Delta J_k (\ar \vec{{\bf T}} \ar) &=& \frac{1}{8\pi^2}\int_0^\infty dq
\frac{q\, n(E_q)\, E_q^{k-1} }{\ar \vec{{\bf T}} \ar} \log \left(
 \frac{2q+\ar \vec{{\bf T}} \ar}{2q-\ar \vec{{\bf T}} \ar} \right)^2,
\qquad k=0,2
 \end{eqnarray*} where
$E_q^2=q^2+m^2$. The above result is valid for real $\ar \vec{{\bf
T}} \ar$ (or $s>4m_\pi^2$ for $\pi\pi$ scattering).

\bibliography{apssamp}

\end{document}